\documentclass[a4paper,12pt]{article}

\usepackage{graphics, epsfig}

\begin{document}

\author{C. Barrab\`es\thanks{E-mail : barrabes@celfi.phys.univ-tours.fr}\\
\small Laboratoire de Math\'ematiques et Physique Th\'eorique,\\
\small CNRS/UMR 6083, Universit\'e F. Rabelais, 37200 TOURS,
France\\P. A. Hogan\thanks{E-mail : phogan@ollamh.ucd.ie}\\ \small
Mathematical Physics Department,\\ \small  National University of
Ireland Dublin, Belfield, Dublin 4, Ireland}

\title{Light--Like Boost of the Kerr Gravitational Field}
\date{}
\maketitle

\begin{abstract}
We describe light--like boosts of the Kerr gravitational field
transverse and parallel to the symmetry axis. In the transverse
case the boosted field is that of an impulsive gravitational wave
having a line singularity displaced relative to its position if
the rotation of the source were removed. The parallel boost is
insensitive to the rotation of the source. The literature contains
a number of diverse results for light--like boosts of the Kerr
gravitational field. Our conclusions confirm the correctness of
the limits calculated by Balasin and Nachbagauer [Class. and
Quantum Grav.{\bf 13}(1996),731]. To avoid any ambiguity our
approach is centered on evaluating the light--like boost of the
Riemann tensor for the Kerr space--time with the metric playing a
secondary role. \vskip 2truepc\noindent To appear in Physical
Review D
\end{abstract}
\vskip 2truepc\noindent PACS number(s): 04.20.Cv
\thispagestyle{empty}
\newpage

\section{Introduction}\indent

Recently \cite{BH} a re--appraisal of the influential work by
Aichelburg and Sexl \cite{AS} on the gravitational field of a
static spherically symmetric body boosted to the speed of light
was carried out. This involved putting the emphasis on the boosted
gravitational field (Riemann tensor) with the metric playing a
secondary role. Working with the Riemann tensor is a guaranteed
way of ensuring that the boosted field is unambiguously determined
because the Riemann tensor is gauge--invariant. This Riemann
tensor centered approach was illustrated by calculating the
boosted gravitational field of a static axially symmetric source.
It is of some importance to apply this new point of view to the
Kerr gravitational field, \emph{particularly in view of the many
differing results which have been derived by authors who have
considered what might be the analogue of the Aichelburg--Sexl
result in this case} \cite{BN1} --\cite{LS2}. Our results are in
general simpler and quite different to those obtained in the
references cited with the notable exception that we find ourselves
in agreement with the limit calculated by Balasin and Nachbagauer
\cite{BN2} (their equation (18) compared to our (2.29) below).
These authors were the first to obtain this transverse light--like
boosted Kerr gravitational field. They introduced a distributional
energy--momentum tensor \cite{BN1} for the Kerr space--time. This
energy--momentum tensor was then subjected to a Lorentz boost and
the light--like limit was taken. The metric of the space--time
after the light--like boost involves a single function which is
obtained by solving a Poisson equation. The procedure is
technically very impressive. However in the light of the existence
of conflicting results in the other works cited, we feel that the
limit calculated in \cite{BN2} could benefit from being confirmed
by the simpler derivation which we supply here.

\setcounter{equation}{0}
\section{Boosted Kerr Field}\indent
We begin with the Kerr line--element \cite{K} in asymptotically
rectangular Cartesian coordinates and time:
\begin{equation}\label{1.1}
ds^2=ds_0^2+\frac{2m\bar r^3}{\bar r^4+a^2\bar z^2}\,\left
[\frac{\bar z}{\bar r}\,d\bar z+\frac{(\bar r\bar x+a\bar y)}{\bar
r^2+a^2}\,d\bar x+\frac{(\bar r\bar y-a\bar x)}{\bar
r^2+a^2}\,d\bar y-d\bar t\right]^2\ ,\end{equation} with
\begin{equation}\label{1.2}
ds_0^2=d\bar x^2+d\bar y^2+d\bar z^2-d\bar t^2\ .
\end{equation}
We use units in which the speed of light $c=1$. Also the constants
$m\,, a$ are the mass and the angular momentum per unit mass
respectively of the source and $\bar r$ is a function of $\bar x,
\bar y, \bar z$ given by
\begin{equation}\label{1.3}
\frac{\bar x^2+\bar y^2}{\bar r^2+a^2}+\frac{\bar z^2}{\bar
r^2}=1\ .
\end{equation}
Thus in coordinates $\bar x^{\mu}=(\bar x, \bar y, \bar z, \bar
t)$ the metric tensor components have the Kerr--Schild form
\cite{KS}\cite{DKS}
\begin{equation}\label{1.4}
\bar g_{\mu\nu}=\bar\eta _{\mu\nu}+2\bar H\,\bar k_{\mu}\bar
k_{\nu}\ ,
\end{equation}
with
\begin{eqnarray}\label{1.5}
\eta _{\mu\nu}\,d\bar x^{\mu}d\bar x^{\nu}&=&d\bar x^2+d\bar
y^2+d\bar z^2-d\bar t^2\ ,\\ \bar k_{\mu}d\bar
x^{\mu}&=&\frac{\bar z}{\bar r}\,d\bar z+\frac{(\bar r\bar x+a\bar
y)}{\bar r^2+a^2}\,d\bar x+\frac{(\bar r\bar y-a\bar x)}{\bar
r^2+a^2}\,d\bar y-d\bar t\ ,\\ \bar H&=&\frac{m\bar r^3}{\bar
r^4+a^2\bar z^2}\ ,
\end{eqnarray}
with the covariant vector field $\bar k_\mu$ null with respect to
$\bar g_{\mu\nu}$ and thus null with respect to the auxiliary
Minkowskian metric tensor $\bar \eta _{\mu\nu}$. Later we shall be
transforming to unbarred coordinates $\{x, y, z, t\}$.

We shall require the coordinate components $\bar
R_{\mu\nu\rho\sigma}$ of the Riemann curvature tensor calculated
from the metric tensor (\ref{1.4}). These are conveniently given
in terms of the complex tensor
\begin{equation}\label{1.8}
{}^+\bar R_{\mu\nu\rho\sigma}=\bar
R_{\mu\nu\rho\sigma}+i\,{}^*\bar R_{\mu\nu\rho\sigma}\ ,
\end{equation}
where the left dual of the Riemann tensor (since the Kerr
space--time is a vacuum space--time the left and right duals of
the Riemann tensor are equal) is defined by ${}^*\bar
R_{\mu\nu\rho\sigma}=\frac{1}{2}\bar \eta
_{\mu\nu\alpha\beta}\,\bar R^{\alpha\beta}{}_{\rho\sigma}$ with
$\bar\eta _{\mu\nu\alpha\beta}=\sqrt{-\bar g}\,\,\epsilon
_{\mu\nu\alpha\beta}$, $\bar g={\rm det}(\bar g_{\mu\nu})$ and
$\epsilon _{\mu\nu\alpha\beta}$ is the Levi--Civita permutation
symbol. We find after a lengthy calculation, using many of the
calculations in \cite{DKS}, that for the Kerr space--time
\begin{equation}\label{1.9}
{}^+\bar R_{\mu\nu\rho\sigma}=-\frac{m\bar r^3}{(\bar r^2+ia\bar
z)^3}\,\left\{\bar g_{\mu\nu\rho\sigma} +i\,\epsilon
_{\mu\nu\rho\sigma}+3\,\bar W_{\mu\nu}\,\bar
W_{\rho\sigma}\right\}\ ,
\end{equation}
with
\begin{equation}\label{1.10}
\bar g_{\mu\nu\rho\sigma}=\bar g_{\mu\rho}\,\bar
g_{\nu\sigma}-\bar g_{\mu\sigma}\,\bar g_{\nu\rho}\ ,
\end{equation}
and the bivector $\bar W_{\mu\nu}=-\bar W_{\nu\mu}$ is given by
the 2--form
\begin{eqnarray}\label{1.11}
\frac{1}{2}\,\bar W_{\mu\nu}\,d\bar x^\mu\wedge d\bar x^\nu
&=&\frac{\bar r}{(\bar r^2+ia\bar z)}\,[\bar x(d\bar x\wedge d\bar
t-i\,d\bar y\wedge d\bar z)+\bar y(d\bar y\wedge d\bar
t\nonumber\\&-&i\,d\bar z\wedge d\bar x)+(\bar z+ia)\,(d\bar
z\wedge d\bar t-i\,d\bar x\wedge d\bar y)]\,.
\end{eqnarray}

We now consider a Lorentz boost in the $-\bar x$ direction with
3--velocity $v<1$. Thus we transform from coordinates $\bar x^\mu
= (\bar x, \bar y, \bar z, \bar t)$ to coordinates $x^\mu =(x, y,
z, t)$ given by
\begin{equation}\label{1.12}
\bar x=\gamma\,(x-vt)\ ,\qquad \bar y=y\ ,\qquad \bar z=z\ ,\qquad
\bar t=\gamma\,(t-vx)\ ,
\end{equation}
with $\gamma =(1-v^2)^{-1/2}$. The components ${}^+\bar
R_{\mu\nu\rho\sigma}$ are transformed to ${}^+
R_{\mu\nu\rho\sigma}$. The transformations naturally divide into
three sets of equations. The first set has $\gamma ^2$ as a factor
on each right hand side and is
\begin{eqnarray}\label{1.13}
{}^+R_{1A1B}&=&\gamma ^2({}^+\bar R_{1A1B}+v\,{}^+\bar
R_{1AB4}+v\,{}^+\bar R_{1BA4}+v^2{}^+\bar R_{A4B4})\ ,\nonumber\\
{}^+R_{1A4B}&=&\gamma ^2({}^+\bar R_{1A4B}+v{}^+\bar
R_{1AB1}+v{}^+\bar R_{4AB4}+v^2{}^+R_{4A1B})\ ,\nonumber\\
{}^+R_{4A4B}&=&\gamma ^2(+\bar R_{4A4B}+v\,{}^+\bar
R_{4AB1}+v{}^+\bar R_{4BA1}+v^2{}^+\bar R_{A1B1})\ ,\nonumber\\
\end{eqnarray}
where the subscripts $A, B$ take values $2, 3$. The second set has
$\gamma$ as a factor on each right side and reads
\begin{eqnarray}\label{1.14}
{}^+R_{1A14}&=&\gamma\,({}^+\bar R_{1A14}+v\,{}^+\bar R_{14A4})\
,\nonumber\\ {}^+R_{14A4}&=&\gamma\,({}^+\bar R_{14A4}+v\,{}^+\bar
R_{1A14})\ ,\nonumber\\ {}^+R_{1ABC}&=&\gamma\,({}^+\bar
R_{1ABC}+v\,{}^+\bar R_{A4BC})\ ,\nonumber\\
{}^+R_{4ABC}&=&\gamma\,({}^+\bar R_{4ABC}+v\,{}^+\bar R_{A1BC})\
.\nonumber\\
\end{eqnarray}
Finally we have the set of transformed components which do not
involve the $\gamma$ factor:
\begin{equation}\label{1.15}
{}^+R_{1414}={}^+\bar R_{1414}\ ,\qquad {}^+R_{1423}={}^+\bar
R_{1423}\ ,\qquad {}^+R_{2323}={}^+\bar R_{2323}\ .
\end{equation}
In the right hand sides of (\ref{1.13})--(\ref{1.15}) the
components ${}^+\bar R_{\mu\nu\rho\sigma}$ are substituted from
(\ref{1.9}) and then the resulting quantities are written in terms
of the coordinates $\{x, y, z, t\}$ using (\ref{1.12}). There is
one relation between these components:
\begin{equation}\label{1.16}
{}^+R_{1234}+{}^+R_{1423}+{}^+R_{1342}=0\ .
\end{equation}

We now in effect ``boost" the Kerr source to the speed of light
transverse to its symmetry axis (the $z$--axis). This is analogous
to the Aichelburg-Sexl \cite{AS} boost of the Schwarzschild field
(to which it specialises when $a=0$) and is achieved by taking the
limit $v\rightarrow 1$ above. The gravitational field of the
boosted source is given by
\begin{equation}\label{1.17}
{}^+\tilde R_{\mu\nu\rho\sigma}=\lim_{v\rightarrow
1}{}^+R_{\mu\nu\rho\sigma}\ .
\end{equation}
\emph{Since the Kerr space--time is a vacuum space--time
its Ricci tensor $\bar R_{\mu\nu}=0$ and hence
the Lorentz transformed Ricci tensor $R_{\mu\nu}=0$ and
therefore $\tilde R_{\mu\nu}=0$.}
In this limit the rest--mass $m\rightarrow 0$ as
$\gamma\rightarrow\infty$ in such a way that the relative energy
$m\,\gamma =p$ (say) remains finite. For a source with multipole
moments, all moments behave like the monopole moment $m$ in the limit
$v\rightarrow 1$ in the case of
the transverse boost (for further explanation of this point see the
calculations described in section IV of \cite{BH}). Thus the parameter $a$
is unaffected by the boost in this case. For a boost parallel to
the symmetry axis the situation, which is quite different, is discussed
in the next section. To evaluate the limit
(\ref{1.17}) explicitly we need the following:
\begin{equation}\label{1.18}
\lim_{v\rightarrow 1}\frac{\gamma\,\bar r^3}{(\bar r^2+ia\bar
z)^3}=\frac{2\,\delta (x-t)}{y^2+(z+ia)^2}\ ,
\end{equation}
and
\begin{equation}\label{1.19}
\lim_{v\rightarrow 1}\frac{\gamma \,\bar r^5}{(\bar r^2+ia\bar
z)^5}=\frac{4}{3}\frac{\delta (x-t)}{(y^2+(z+ia)^2)^2}\ ,
\end{equation}
where $\delta (x-t)$ is the Dirac delta function singular on
$x=t$. To establish (\ref{1.18}) we note that (\ref{1.3}) and
(\ref{1.12}) can be used to write
\begin{equation}\label{1.20}
\frac{\gamma\,\bar r^3}{(\bar r^2+ia\bar
z)^3}=\frac{1}{(y^2+(z+ia)^2)}\,\frac{\partial}{\partial x}\left (
\frac{(x-vt)\,\bar R}{\bar R^2+i\gamma ^{-2}az}\right )\ .
\end{equation}
The quantity $\bar R$ is defined as follows: first we have from
(\ref{1.3}) that
\begin{equation}\label{1.21}
\bar r^2=\frac{1}{2}\{r^2-a^2+\sqrt{(r^2-a^2)^2+4a^2z^2}\}\ ,
\end{equation}\label{1.22}
with $r^2=\bar x^2+\bar y^2+\bar z^2=\gamma ^2(x-vt)^2+y^2+z^2$.
Thus we have
\begin{equation}\label{1.23}
r=\gamma\,R\ ,\qquad R=\sqrt{(x-vt)^2+\gamma ^{-2}(y^2+z^2)}\ ,
\end{equation}
and so (\ref{1.21}) yields
\begin{equation}\label{1.24}
\bar r=\gamma\,\bar R\ ,\qquad \bar R^2=\frac{1}{2}\{R^2-\gamma
^{-2}a^2+\sqrt{(R^2-\gamma ^{-2}a^2)^2+4\gamma ^{-4}a^2z^2}\}\ ,
\end{equation}
giving us $\bar R$. With $\bar R$ and $R$ now defined we see from
(\ref{1.20}) that
\begin{equation}\label{1.25}
\lim_{v\rightarrow 1}\frac{\gamma\,\bar r^3}{(\bar r^2+ia\bar
z)^3}=\frac{1}{(y^2+(z+ia)^2)}\,\frac{\partial}{\partial x} \left
(\frac{x-t}{|x-t|}\right )\ .
\end{equation}
If $\theta (x-t)$ is the Heaviside step function which is equal to
unity if $x>t$ and is zero if $x<t$ then
\begin{equation}\label{1.26'}
\frac{x-t}{|x-t|}=2\,\theta (x-t)-1\ ,
\end{equation}
and since $\partial\theta /\partial x=\delta (x-t)$ we recover
(\ref{1.18}) from (\ref{1.25}).

By (\ref{1.3}) and (\ref{1.12}) we have
\begin{equation}\label{1.26}
\frac{\partial\bar r}{\partial y}=\frac{y\,\bar r^3}{\bar
r^4+a^2\bar z^2}\ ,
\end{equation}
and using this we find that (\ref{1.19}) follows from (\ref{1.18})
by differentiating (\ref{1.18}) with respect to $y$ and taking
$y\neq 0$. When $a=0$ we note that (\ref{1.18}) and (\ref{1.19})
reduce to (2.10) and (3.5) respectively of \cite{BH}

We now evaluate the limits (\ref{1.17}). As an illustration we
find
\begin{equation}\label{1.27}
{}^+\tilde R_{1212}=\lim_{v\rightarrow 1}\left [\frac{3m\gamma
^2\bar r^5}{(\bar r^2+ia\bar z)^5}\,\{z+i(a+y)\}^2\right ]=
4p\,\left [\frac{z+i(a+y)}{y^2+(z+ia)^2}\right ]^2\delta (x-t)\ ,
\end{equation}
by (\ref{1.19}). After some algebra this can be rewritten as
\begin{equation}\label{1.28}
{}^+\tilde R_{1212}=4p\left [\frac{z+i(y-a)}{z^2+(y-a)^2}\right
]^2\delta (x-t)\ ,
\end{equation}
and if
\begin{equation}\label{1.29}
H=2p\,\log\{(y-a)^2+z^2\}\,\delta (x-t)\ ,
\end{equation}
we finally have
\begin{equation}\label{1.30}
{}^+\tilde R_{1212}=\left (H_{yy}-iH_{yz}\right )\ .
\end{equation}
The subscripts on $H$ denote second partial derivatives. In
similar fashion we find that $\tilde R_{\mu\nu\rho\sigma}\equiv 0$
except for
\begin{eqnarray}\label{1.31}
{}^+\tilde R_{1212}&=&{}^+\tilde R_{2424}=-{}^+\tilde
R_{1313}=-{}^+\tilde R_{3434}={}^+\tilde R_{3134}=-{}^+\tilde
R_{2124} \nonumber\\ &=&\left (H_{yy}-iH_{yz}\right )\ ,
\end{eqnarray}
and
\begin{eqnarray}\label{1.32}
{}^+\tilde R_{1213}&=&{}^+\tilde R_{2434}=-{}^+\tilde
R_{3124}=-{}^+\tilde R_{2134} \nonumber\\
&=&i\left (H_{yy}-iH_{yz}\right )\ ,
\end{eqnarray}

When the Lorentz transformation (\ref{1.12}) is applied to the
line--element (\ref{1.1}) and the limit $v\rightarrow 1$ taken we
find that \emph{if $x-t>0$ then}
\begin{equation}\label{1.33}
\lim_{v\rightarrow
1}ds^2=dx^2+dy^2+dz^2-dt^2+\frac{8\,p}{x-t}\,(dx-dt)^2\ ,
\end{equation}
and \emph{if $x-t<0$ then}
\begin{equation}\label{1.34}
\lim_{v\rightarrow 1}ds^2=dx^2+dy^2+dz^2-dt^2\ .
\end{equation}
For $x-t>0$ we can write (\ref{1.33}) in the form
\begin{equation}\label{1.35}
ds_+^2=dy_+^2+dz_+^2+2\,du\,dv_+\ ,
\end{equation}
with
\begin{eqnarray}\label{1.36}
y_+&=&y\,\qquad z_+=z\ ,\qquad u=x-t\ ,\nonumber\\
v_+&=&\frac{1}{2}\,(x+t)+4\,p\,\log (x-t)\ .
\end{eqnarray}
For $x-t<0$ we can write (\ref{1.34}) in the form
\begin{equation}\label{1.37}
ds_-^2=dy_-^2+dz_-^2+2\,du\,dv_-\ ,
\end{equation}
with
\begin{eqnarray}\label{1.38}
y_-&=&y\,\qquad z_-=z\ ,\qquad u=x-t\ ,\nonumber\\
v_-&=&\frac{1}{2}\,(x+t)\ .
\end{eqnarray}
From (\ref{1.35}) and (\ref{1.37}) we see that $u=0$ is a null
hyper{\it plane} in Minkowskian space--time. The line--elements
(\ref{1.35}) and (\ref{1.37}) are consistent with having a delta
function in the Riemann curvature tensor which is singular on
$x=t$ provided the two halves of Minkowskian space--time , $x>t$
and $x<t$, are attached on $x=t$ with
\begin{equation}\label{1.39}
y_+=y_-\ ,\qquad z_+=z_-\ ,\qquad v_+=F(v_-, y_-, z_-)\ ,
\end{equation}
for some function $F$ defined on $x=t$ for which $\partial F/
\partial v_-\neq 0$ \cite{BH2}. The particular function $F$ which
gives rise to the coefficients of the delta function in the
Riemann tensor components listed in (\ref{1.31}) and (\ref{1.32})
can be calculated from the formulas given in \cite{BH} (and
originally derived in \cite{BH2}). It is easily found to be
\begin{equation}\label{1.40}
v_+=F=v_-+2\,p\,\log\{(y-a)^2+z^2\}\ .
\end{equation}
The signal with history $x=t$ is an impulsive gravitational wave
with a delta function profile and which is, in addition, singular
on the null geodesic generator $y=a\ ,z=0$ of the null hyperplane
$x=t$. This is identical to the light--like boosted Schwarzschild
field \cite{AS} except that the singular generator is shifted from
$y=0\ ,z=0$ in that case to $y=a\ ,z=0$. The physical significance
of this shifted generator (the imprint in the boosted field of the
rotation of the source of the original Kerr gravitational field)
can be elucidated, for example, by studying the deflection of
highly relativistic particles in the Kerr gravitational field
\cite{Def}.

Once $F$ in (\ref{1.40}) is known the line--element of the
space--time can be constructed in coordinates in which the metric
tensor is continuous across $x=t$ \cite{BH2}. For the light--like
boosted Kerr field the resulting line--element can then be
transformed into Kerr--Schild form at the expense of introducing
the delta function into the metric tensor. This line--element is
\begin{equation}\label{1.40'}
ds^2=dx^2+dy^2+dz^2-dt^2-2\,H\,(dx-dt)^2\ ,
\end{equation}
with $H$ given by (\ref{1.29}). The corresponding Riemann tensor
(\ref{1.31}) and (\ref{1.32}) is type N in the Petrov classification
with degenerate principal null direction given via the 1--form
$dx-dt$.

\setcounter{equation}{0}
\section{Discussion}\indent
The light--like boost of the Kerr gravitational field described
above involves a boost \emph{transverse} to the symmetry axis of
the Kerr source. To complete the picture we consider briefly here
the simpler case of a light--like boost in the $-\bar
z$--direction which is parallel to the symmetry axis. To motivate
our approach to this we expand $F$ given in (\ref{1.40}) for the
transverse light--like boost in powers of the parameter $a$.  Thus
we arrive at
\begin{equation}\label{1.41}
F=v_-+2\,\sum_{l=0}^{\infty}(-1)^l\,\frac{p_l}{l!}\,
\frac{\partial ^l}{\partial y^l}\left (\log (y^2+z^2)\right )\ ,
\end{equation}
with $p_l=p\,a^l$ for $l=0, 1, 2, \dots $. This is similar to the
matching function $F$ encountered in the transverse light--like
boost of a static axially symmetric multipole field \cite{BH}
whose original multipole moments, before taking the limit
$v\rightarrow 1$, were
\begin{equation}\label{1.42}
A_l=m\,a^l\ ,\qquad l=0, 1, 2, \dots\ ,
\end{equation}
and with $m=p\,\gamma ^{-1}$. For the light--like boost in the
$-\bar z$--direction (the direction of the symmetry axis) the
analogy with the multipole field \cite{BH} suggests that we take
$a=\gamma ^{-1}\hat a$ as $v\rightarrow 1$ in this case, aswell as
taking $m=p\,\gamma ^{-1}$. Then (\ref{1.42}) is replaced by
\begin{equation}\label{1.43}
A_l=\gamma ^{-l-1}p_l\ ,\qquad p_l=p\,\hat a^l\ ,
\end{equation}
for $l=0, 1, 2,\dots $ as in the case of a light--like boost
parallel to the symmetry axis in \cite{BH}. Now make the Lorentz
transformation
\begin{equation}\label{1.44}
\bar x=x\ ,\qquad \bar y=y\ ,\qquad \bar z=\gamma\,(z-vt)\ ,\qquad
\bar t=\gamma\,(t-vz)\ .
\end{equation}
The components ${}^+\bar R_{\mu\nu\rho\sigma}$ in (\ref{1.9}) are
transformed to ${}^+R_{\mu\nu\rho\sigma}$. In this case the
equations (\ref{1.13})and (\ref{1.14}) are replaced by the same
equations, with the subscript $1$ replaced by the subscript $3$
and with $A, B, C$ each taking the values $1, 2$. Also the
subscripts $1$ and $3$ are interchanged in equations (\ref{1.15}).
The relation (\ref{1.16}) continues to hold. Now (\ref{1.19}) is
replaced by
\begin{equation}\label{1.45}
\lim_{v\rightarrow 1}\frac{\gamma\,\bar r^5}{(\bar r^2+ia\bar
z)^5}=\lim_{v\rightarrow 1}\frac{\gamma\,\bar r^5}{\{\bar
r^2+i\hat a(z-vt)\}^5}=\lim_{v\rightarrow 1}\frac{\gamma
^{-4}}{\tilde R^5}=\frac{4}{3}\,\frac{\delta (z-t)}{(x^2+y^2)^2}\
,
\end{equation}
where $\tilde R=\sqrt{(z-vt)^2+\gamma ^{-2}(x^2+y^2)}$. The last
equality in (\ref{1.45}) is derived as Eq.(3.5) in \cite{BH}.
Calculating ${}^+\tilde R_{\mu\nu\rho\sigma}$ in (\ref{1.17}) in
this case we find that the non--identically vanishing components
are
\begin{eqnarray}\label{1.46}
{}^+\tilde R_{2323}&=&{}^+\tilde R_{2424}=-{}^+\tilde
R_{2324}=-{}^+\tilde R_{3131}=-{}^+\tilde R_{1414}={}^+\tilde
R_{1314}\nonumber\\ &=&\frac{4\,p\,(x-iy)^2}{(x^2+y^2)^2}\,\delta
(z-t)\ ,
\end{eqnarray}
and
\begin{equation}\label{1.47}
{}^+\tilde R_{3231}={}^+\tilde R_{2414}=-{}^+\tilde
R_{1324}=-{}^+\tilde
R_{2314}=-\frac{4ip\,(x-iy)^2}{(x^2+y^2)^2}\,\delta (z-t)\ .
\end{equation}
This is the same field as the light--like boosted Schwarzschild
field derived by Aichelburg and Sexl \cite{AS}. The generalisation
of the calculations given here to a light--like boost of the Kerr
gravitational field in an arbitrary direction involves tedious
algebra but is straightforward and would result in an explicit
limit for the Riemann tensor.


\begin{thebibliography}{99}
\bibitem{BH} C. Barrab\`es and P. A. Hogan, Phys. Rev. D{\bf 64},
044022 (2001).
\bibitem{AS} P. C. Aichelburg and R. U. Sexl, Gen. Relativ.
Gravit. {\bf 12}, 303 (1971).
\bibitem{BN1} H. Balasin and H. Nachbagauer, Class. Quantum Grav.
{\bf 12}, 707 (1995).
\bibitem{BN2} H. Balasin and H. Nachbagauer, Class. Quantum Grav.
{\bf 13}, 731 (1996).
\bibitem{BM} A. Burinskii and G. Magli, Phys. Rev. D{\bf 61}, 044017 (2000).
\bibitem{FP} V. Ferrari and P. Pendenza, Gen. Relativ. Gravit.
{\bf 22}, 1105 (1990).
\bibitem{LS1} C. O. Lousto and N. Sanchez, Nucl. Phys. B{\bf 355},
231 (1991).
\bibitem{LS2} C. O. Lousto and N. Sanchez, Nucl. Phys. B{\bf 383},
377 (1992).
\bibitem{K} R. P. Kerr, Phys. Rev. Letters {\bf 11},
237 (1963).
\bibitem{KS} R. P. Kerr and A. Schild, \emph{Atti del Convegno sulla Relativit\`a Generale:
Problemi dell Energia e Onde Gravitazionali (Anniversary Volume,
Fourth Centenary of Galileo's Birth),} G. Barb\'era, Ed. (Firenze,
1965), p. 173; \emph{Applications of Nonlinear Partial
Differential Equations in Mathematical Physics, Proceedings of
Symposia in Applied Mathematics} (American Mathematical Society,
Providence, R.I., 1965), vol. XVII, p. 199.
\bibitem{DKS} G. C. Debney, R. P. Kerr and A. Schild, J. Math. Phys. {\bf 10}, 1842
(1969).
\bibitem{Def} C. Barrab\`es and P. A. Hogan, ``The Deflection of
Highly Relativistic Particles in a Gravitational Field" (preprint)
2002.
\bibitem{BH2} C. Barrab\`es and P. A. Hogan, Int. J. Mod. Phys.
D{\bf 10}, 711 (2001).

\end{thebibliography}
\end{document}